\documentclass[aps,pra,showpacs,superscriptaddress,preprint]{revtex4}
\usepackage{epsf}
\usepackage{amsmath}
\usepackage{epsfig}

\begin{document}

\title{Solvation Effects on Free Energy Surface of Polyalanine}

\author{G\"{o}khan G\"{o}ko\u{g}lu}
\email[E-mail: ]{ggokoglu@hacettepe.edu.tr} \affiliation{Hacettepe
University, Department of Physics Engineering , Beytepe, 06800
Ankara, Turkey}
\author{Handan Arkin Ol\u{g}ar}
\email[E-mail: ]{handan@hacettepe.edu.tr} \affiliation{Hacettepe
University, Department of Physics Engineering , Beytepe, 06800
Ankara, Turkey}
\author{Ethem Akt\"urk}
\email[E-mail: ]{eakturk@hacettepe.edu.tr} \affiliation{Hacettepe
University, Department of Physics Engineering , Beytepe, 06800
Ankara, Turkey}
\author{Tarik \c{C}elik}
\email[E-mail: ]{tcelik@hacettepe.edu.tr}\affiliation{Hacettepe
University, Department of Physics Engineering , Beytepe, 06800
Ankara, Turkey}
\begin{abstract}
We have simulated
10-residue polyalanine chain by multicanonical method to visualize
the 3D topographic picture of the free energy landscape over the whole
 range of temperatures,
hence to show the funnel along the folding pathway  exhaustively.
We have simulated and compared the system in vacuo and in solvent,
and examined the changes in the free energy landscape due to the
solvent effects, which are taken into account by commonly used
model Accessible Surface Area.

\noindent Keywords: Free energy funnel, solvation effects,
polyalanine. \vspace{4mm} \pacs{02.70.Uu Applications of Monte
Carlo Methods, 05.10.-a Computational Methods in Statistical
Physics and Nonlinear Dynamics, 82.20.Wt Computational Modelling
Simulation.}

\vspace{4mm}

\end{abstract}
\maketitle
\newpage
\section{Introduction}

~~~The configuration space of proteins presents a complex
energy profile consisting of tremendous number of local minima,
barriers, attraction basins and further topological features.
The problem of protein folding entails the study of a non-trivial
dynamics along pathways embedded in a rugged energy landscape.
The topology of the landscape
characterizes the folding kinetics through the existence of folding pathway.

\bigskip

The topography of the energy landscape,
especially near the global minimum, is of particular importance, because
the potential energy surface defines the behavior of the system~\cite{JBPW}.
However, the entropic effects play a major role in the folding process, and
instead of potential energy, the landscape of the free energy surface
better be examined. The essence of a funnel structure of free energy at
some fixed temperatures has been shown by Hansmann and
Onuchic~\cite{HaOn99}.
A visualization of the whole rugged free energy landscape covering the
entire temperature range would lead to clear
indications of the equilibrium conformations of proteins as a function of
the temperature and provide a detailed picture of the folding pathway.

\bigskip

On the other hand, it is vital in simulations to mimic the effects
of the environment
which change the topology of the free energy surface to a large extent,
hence alter the phase transition behaviour of the system. The solution
effects can be included by incorporating the Accessible Surface Area (ASA),
in which model
the solvation energy term is proportional to accessible surface area of the
peptide.

\bigskip

In this study, our aim is to obtain a 3D topographic picture of free energy
surface covering the whole temperature range, display the funnel along the
folding pathway and examine how the solution
effects alter the free energy landscape.
Such a goal can be achieved within the multicanonical ensemble
approach~\cite{HATC}.

\bigskip

The conventional
simulation methods are not effective because
the system becomes trapped for long simulation time in a potential
well.
The trapping problem of the Monte Carlo and Molecular Dynamics
 methods can
be alleviated to a large extent, by the Multicanonical MC
method (MUCA)~\cite{BeCe92,Be99}, which  was applied initially
to lattice spin models and its relevance for complex systems was first
noticed in
Ref.~\cite{BeCe92}.  Application of the multicanonical approach
  to peptides was pioneered by
Hansmann and Okamoto~\cite{HaOk93} and followed by
others~\cite{HaSc94}; simulations of protein folding with
MUCA and related generalized ensemble methods are reviewed in
Refs. ~\cite{HaOk99rev} and ~\cite{Ok00rev}.

\bigskip

The multicanonical ensemble  based on a probability
function in which
the different energies are equally probable.
However, implementation of  MUCA is not straightforward
because the density of states  $n(E)$ is unknown {\it a priori}.
The weights $ w(E) \sim 1/n(E)$ are calculated in the
first stage of simulation  process  by an iterative  procedure.
The iterative  procedure is
followed by a long production run based on the fixed $w$'s where
equilibrium configurations are sampled. Re-weighting techniques
(see Ferrenberg and Swendsen~\cite{FeSw88} and literature given in
their second reference)
enable one to obtain Boltzmann averages of various
thermodynamic properties over a large range of temperatures.

\bigskip

By setting up a one-dimensional random walk in energy space,
the multicanonical simulation provides sampling of all available energies
and enables one to study the thermodynamical aspects of the system at
a wide range of temperature from a single production run. In this work, we will
exploit the multicanonical ensemble to investigate and compare the free
energy landscape of the 10-residue polyalanine chain in vacuo and in solvent.


\section{Methods}

ECEPP (Empirical Conformational Energies for Proteins and Polipeptides)
force field, one of the most commonly used all atom force field was
used in simulations. This force field composed of four potential energy terms;
electrostatic, hydrogen bond, Lennard-Jones and torsion energy term.

\begin{eqnarray}
E_{ECEPP/2} & = & E_{C} + E_{LJ} + E_{HB} + E_{tor},\\
E_{C}  & = & \sum_{(i,j)} \frac{332q_i q_j}{\epsilon r_{ij}},\\
E_{LJ} & = & \sum_{(i,j)} \left( \frac{A_{ij}}{r^{12}_{ij}}
                                - \frac{B_{ij}}{r^6_{ij}} \right),\\
E_{HB}  & = & \sum_{(i,j)} \left( \frac{C_{ij}}{r^{12}_{ij}}
                                - \frac{D_{ij}}{r^{10}_{ij}} \right),\\
E_{tor}& = & \sum_l U_l \left( 1 \pm \cos (n_l \chi_l ) \right).
\label{ECEPP/2}
\end{eqnarray}

Here, $r_{ij}$ (in \AA) is the distance between the atoms $i$ and $j$,
$\chi_l$ is the torsion angle for the chemical bond $l$ and $n_l$
characterizes its symmetry. The force field parameters $q_i,A_{ij},B_{ij},
C_{ij}, D_{ij},U_l$ were calculated from crystal structures.  The dielectricity
constant is set to $\varepsilon=2$, its common value in ECEPP calculations.
 ECEPP
force field is based on rigid geometry that means bond lengths and bond
 angles are
 constant. The backbone torsion angles $\phi$, $\psi$ and the side
chain torsion angle $\chi$ are the degrees of freedom of the system.
$\omega$ torsion angle was fixed in trans position that $\omega$=$180^{\circ}$.
Solvation energy term is~\cite{Ooi},

\begin{equation}
E_{solv}=\sum_i\sigma_iA_i~,
\end{equation}
where  $A_i$ is the solvent accessible surface area
of the $i^{th}$ atom for a definite conformation, and $\sigma_i$
the  solvation parameter for the atom $i$. The total potential energy of
the molecule
is then given by

\begin{equation}
E_{tot} = E_{ECEPP} + E_{solv}
\end{equation}

This potential energy is implemented
into the software package FANTOM~\cite{FANTOM}.

\bigskip

The order parameter chosen for the system is the normalized helicity $q$, which
is defined as~\cite{HaOk99},

\begin{equation}
q = \frac{n_H}{N-2}
\end{equation}
where $n_H$ is the number of residues that conforms to $\alpha$-helix
state defined in the angle range for $\phi$ $(-70 \pm 20^{\circ})$ and
for $\psi$ $(-37 \pm 20^{\circ})$, $N$ is the total length of the chain.
End residues are more flexible than others and was not taken into account.
This definition guarantees the q value
between $0$ and $1$.

\bigskip

Helmholtz free energy of the system was calculated by the formula
\begin{equation}
F(q,T) = - RT\log P(q)
\end{equation}
where $P(q)$ is the probability that the system has an order parameter value
$q$ at a fixed temperature $T$ and $R$ is the gas constant.
This free energy is not
normalized to be expressed in unit kcal/mol, however it gives
a qualitative view of free energy surface.

\bigskip

We first carried out canonical (i.e.,
constant $T$) MC simulations at relatively high temperatures and
MUCA test runs which enabled us to determine the required energy
ranges. The energy range was divided into
40 bins of $1\,$kcal/mol each, covering the range $[-10,30]\,$kcal/mol for vacuo
and 25 bins of $1\,$kcal/mol each, covering the range $[-23,2]\,$kcal/mol for
solvent. At each update step, a
trial conformation was obtained by changing {\it one} dihedral angle
at random within  the range [$-180^{\rm o};180^{\rm o}]$, followed by
the Metropolis test and an update of the suitable histogram. The
dihedral angles were always visited in a predefined (sequential)
order; a cycle of $N$ MC steps ($N$=30) is
called a sweep.
The weights  were built after $ 150 $ recursions during a
long {\it single} simulation, where the multicanonic parameters
were iterated every 5000~sweeps.
Then we performed full simulation of two million sweeps, which cover the
temperature region up to $T_{\max}=600\,$K reliably.

\section{Results and Discussion}

~~~Polyalanine is an important model system to study specially the
backbone behaviour in polypeptide systems with small side chains,
so that a great deal of theoretical works focus on it. The major goal of these
studies is to determine the stable conformation and to investigate
the structural properties of the system. The success of simulation
studies strongly depends how well the nature environment
of the polypeptide system was described. Hence the solvation effects
must be included in simulation otherwise the results does not reflect
the physical reality.

\bigskip

Free energy calculations in polypeptide systems in a temperature range
is not straightforward
and generally the potential energy is considered. However, the
stable conformation of the system corresponds to the global minimum of the
free energy rather than the potential energy, so that entropic
contributions are taken into account.

\bigskip

By utilizing the multicanonical technique,
we have obtained 3D topographic picture of
the free energy surface for
10-residue polyalanine chain in vacuo over whole temperatures and plotted vs.
the order parameter in Fig.1a.
It can be seen from the figure that there is no data for some $q(T)$
values, which means no histograms detected in that region and these states
can be regarded as non-accessible.
The free energy surface displays an apparent valley structure, which
clearly pictures the existing funnel towards the state of global energy
minimum (GEM).
Free energy surface has two wings divided
by a valley along which the free energy is lower than other parts of the
surface. The states with lower free energy lie in the riverbed along the
valley and serves one to visualize the folding pathway.
From high temperatures to
lower ones, the system will follow that valley from one corner of the
$q-T$ plane to other and
will make a transition from a disordered state to highly ordered state that
is the well known helix-coil transition. The ordered state $(q=1)$ has the minimum
free energy  which is the stable conformation for the systems under study.
Global energy minima found in our simulations of the system in vacuo
and in solvent are
$E=-9.9722$ kcal/mol and $E=-22.8079$ kcal/mol, respectively, both conform
to a stable $\alpha$-helix structure.
Our results supports that of Scholtz et al. suggesting that longer polyalanine
chains form stable $\alpha$-helices in water~\cite{Scholtz}. The system
 in vacuo has a big tendency to
$\alpha$- helical conformations that the system is in exact helical state up
to temperatures $400 K$. It is known that polyalanine has an intrinsic
tendency to $\alpha$- helical conformations.

\bigskip

The free energy surface of the 10-residue polyalanine chain in solvent is
displayed in Fig.1b.
If we compare the free energy surfaces of the same system in vacuo
and in solvent, the valley is quite distorted, one of the wings is almost
disappeared and the whole picture is pulled down to lower temperatures.
In both figures, the side with lower free energy is
more probable hence the system with solvent is in highly disordered state at
temperatures higher than $150 K$.
The disordered states are still probable in solvent
with low $q$ value at temperatures as low as $100K$.

\bigskip

Drastic changes in free energy
surface by putting the system in solvent can be attributed to that the
solvent used in simulations is water which is a good solvent specially for
polyalanine chain. Due to large increase in entropy, free energy
surface becomes much altered in solvent. This massive effect can also
be seen in
Fig. 2, where the phase transition temperature are  $490 K$ and
$170 K$ for the systems vacuo and solvent, respectively. Decrease in specific
heat for the system in solvent is an expected result for peptides.

\bigskip

In Fig.3, free energy is shown as a function of the order parameter
at distinct temperatures for the system in vacuo. At critical temperature
$T_c=490 K $
free energy makes a deeper valley than other temperatures and the whole
sequence of fixed temperature plots mimics a smooth phase transition.
But it must be remembered that, in
small systems, a first order phase transition may be seen as if it is
continuous.

\section{Conclusion}

In conclusion, we have simulated the 10-residue polyalanine
chain by utilizing multicanonical ensemble approach and
investigated the structure of
rugged free energy landscape in configurational space.
We were able to display the distribution of all conformations in the
configuration space
at all temperatures from a single simulation. From the topographic
picture obtained, one can visualize the structure of the folding pathway
and the changes due to solution effects.
Such a visualization of the rugged
energy landscape would certainly be helpful in another aspect, namely
in designing algorithms for efficient sampling of conformational space.

\bigskip

\noindent
{\bf Acknowledgments}: \\
This work is supported by Hacettepe University Scientific Research
Fund through project No: 02.02.602.010.



\vfil

\clearpage
\newpage
{\huge Figure Captions:}
\begin{description}
\item[Fig.~1] Free energy surface of polyalanine in vacuo
(upper)and in solvent (lower). The funnel structure is
exhaustively displayed. \item[Fig.~2] Specific heat as a function
of the temperature for 10-residue
              polyalanine chain in vacuo and in solvent.
\item[Fig.~3] Free energy as a function of the order parameter at $T=380,
              460, 490$ and $520 K$.
\end{description}

\begin{figure}[htbp]
\begin{center}
\setlength{\unitlength}{1cm} \epsfig{file=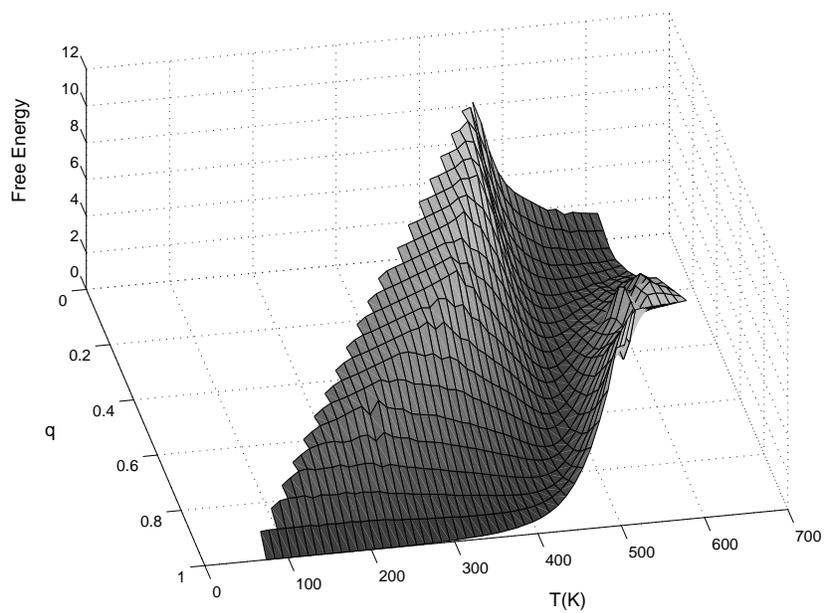,width=11cm}
\epsfig{file=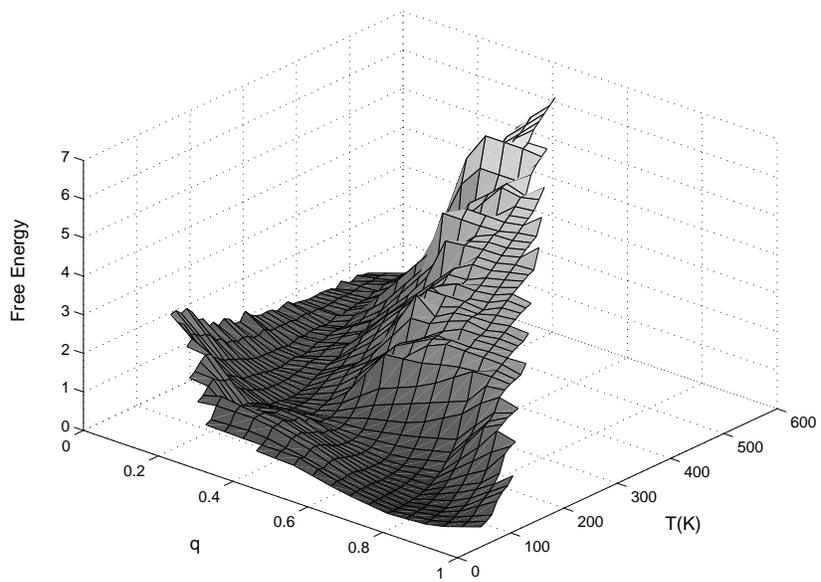,width=11cm} \caption{Free energy surface
of polyalanine in vacuo (upper) and in solvent (lower). The funnel
structure is exhaustively displayed.}
\endgroup
\end{figure}

\clearpage
\newpage
\begin{figure}[htbp]
\setlength{\unitlength}{1cm} \epsfig{file=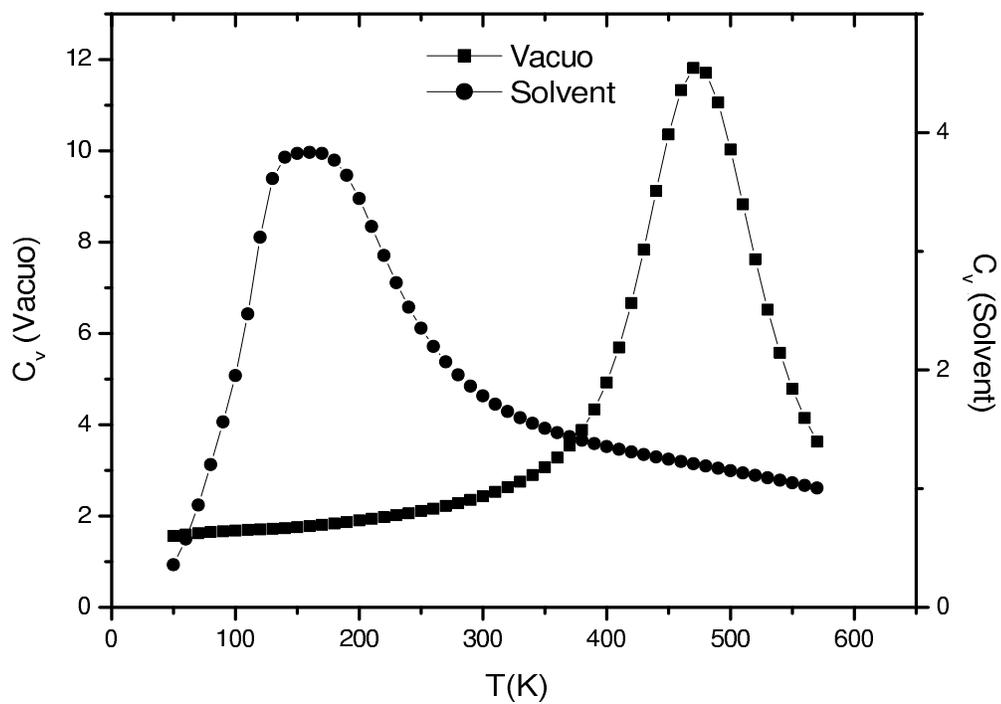,width=14cm}
\caption{Specific heat as a function of the temperature for
10-residue polyalanine chain in vacuo and in solvent.}
\end{figure}

\clearpage
\newpage
\begin{figure}[htbp]
\setlength{\unitlength}{1cm} \centerline{\hbox{
\epsfig{file=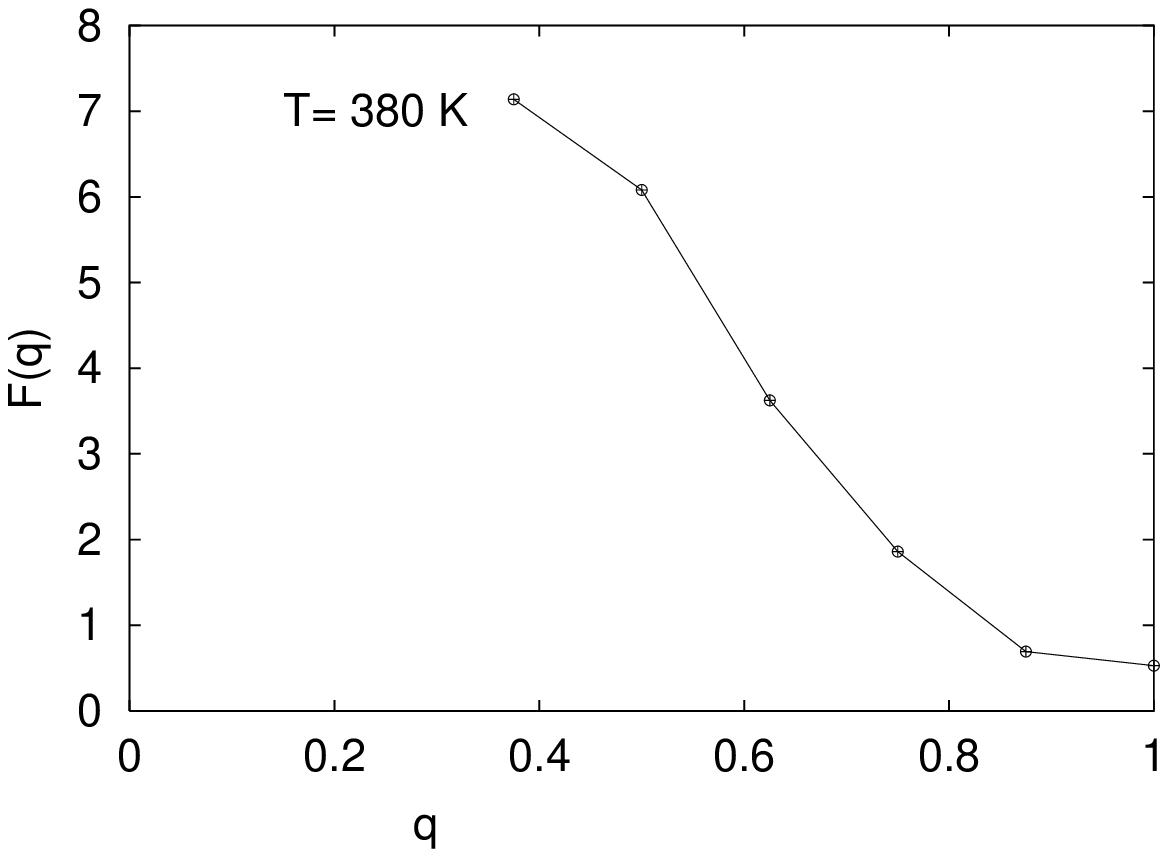,width=7cm}
\epsfig{file=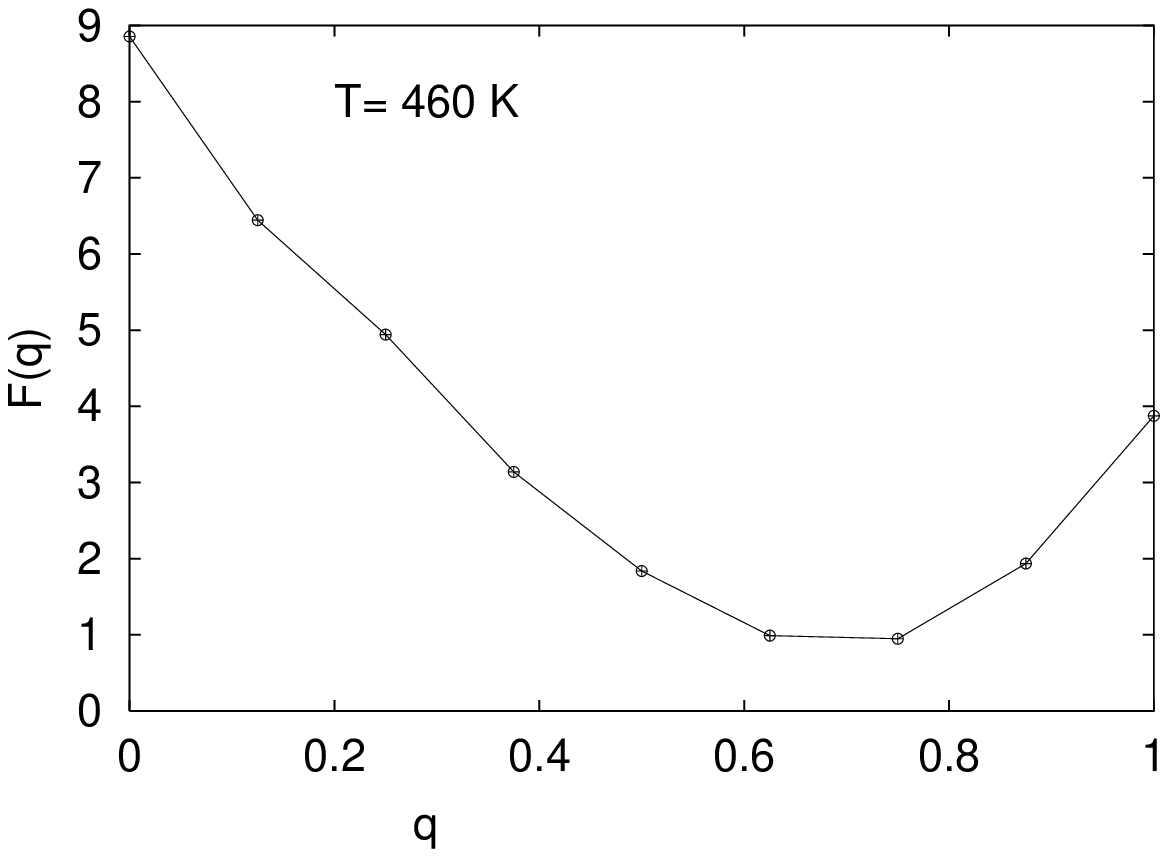,width=7cm}
 }}
\vspace*{-0.5cm}
\end{figure}
\begin{figure}[htbp]
\centerline{\hbox{ \psfig{file=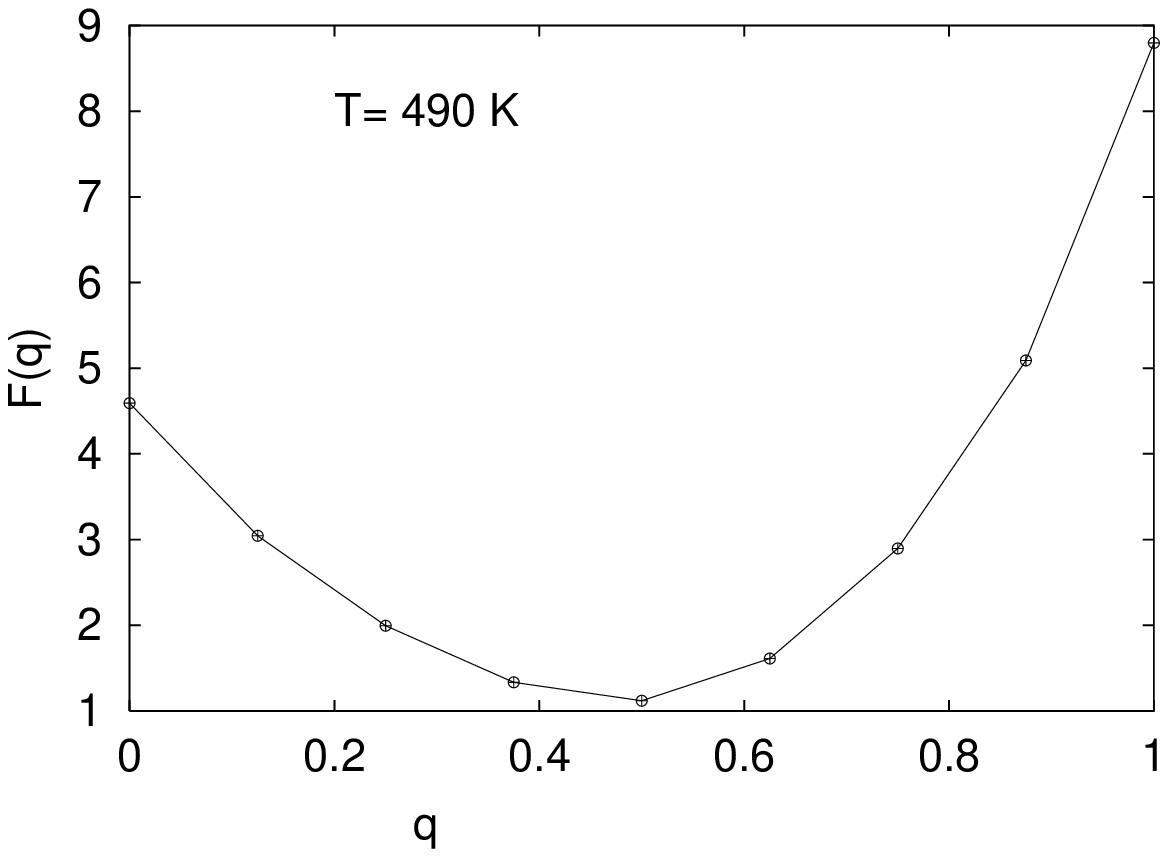,width=7cm}
\psfig{file=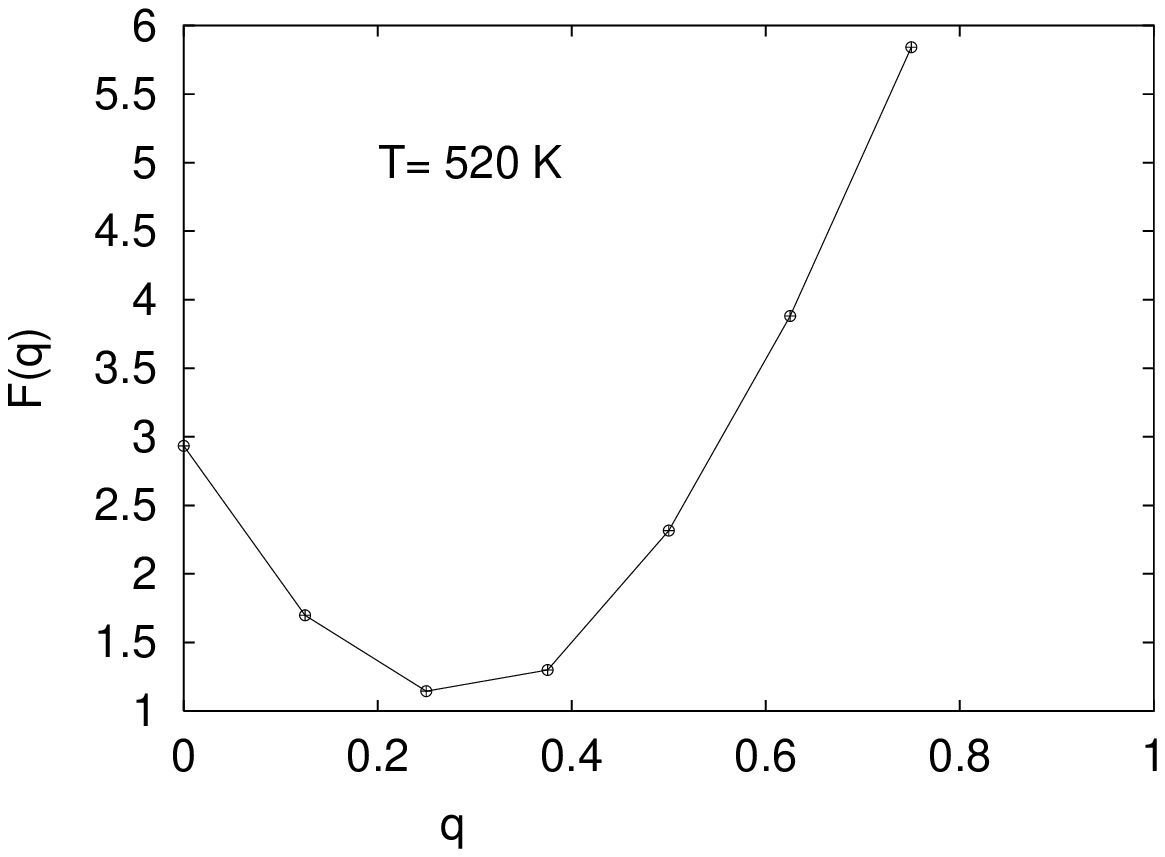,width=7cm} }} \caption{Free energy as a
function of the order parameter at $T=380, 460, 490$ and $520 K.$}
\end{figure}


\begin{thebibliography}{99}

\bibitem{JBPW}   J.D. Bryngelson, P.G. Wolynes,
                  {\it Proc. Natl. Acad. Sci.} {\bf 84}, 7524 (1987).

\bibitem{HaOn99}  U.H.E. Hansmann, Y. Okamoto, J.N. Onuchic,
                     {\it Proteins} {\bf 34}, 472 (1999) ;
                   U.H.E. Hansmann, J.N. Onuchic,
                    {\it J. Chem. Phys.} {\bf 115}, 1601 (2001).

\bibitem{HATC}   H. Ark{\i}n, T. \c{C}elik,
                  {\it Int. J. Mod. Phys. C} {\bf 14}, 113 (2003).

\bibitem{BeCe92}  B.A. Berg, T. \c{C}elik,
                     {\it Phys. Rev. Lett.} {\bf 69}, 2292 (1992).

\bibitem{Be99}    B.A. Berg,
                    {\it Fields Institute Communications} {\bf 28}, 1 (1992).

\bibitem{HaOk93}  U.H.E. Hansmann, Y. Okamoto,
                   {\it J. Comp. Chem.} {\bf 14}, 1333 (1993).


\bibitem{HaSc94}  M.H. Hao, H.A. Scheraga,
                  {\it J. Phys. Chem.} {\bf 98}, 4940 (1994) ;
                   {\it J. Phys. Chem.} {\bf 98}, 9882 (1994);
                   A. Kolinski, W. Galazka, J. Skolnick,
                    {\it Proteins} {\bf 26}, 271 (1996);
                   J. Higo, N. Nakajima, H. Shirai, A. Kidera,
                   H. Nakamura, {\it J. Comp. Chem.} {\bf 18}, 2086 (1997).


\bibitem{HaOk99rev} U.H.E. Hansmann, Y. Okamoto,
                    {\it Ann. Rev. Comp. Physics} {\bf 5}, 129 (1999).

\bibitem{Ok00rev} A. Mitsutake, Y.  Sugita, Y. Okamoto,
                   {\it Biopolymers (Peptide Science)} {\bf 60}, 96 (2001).

\bibitem{FeSw88}  A.M. Ferrenberg, R.H. Swendsen,
                  {\it Phys. Rev. Lett.} {\bf 61}, 2635 (1988);
                  {\it Ibid} {\bf 63}, 1658 (1989).

\bibitem{Ooi} T. Ooi, M. Obatake, G. Nemethy, H.A. Scheraga,
                   {\it Proc. Natl. Acad. Sci.} {\bf 8}, 3086 (1987).

\bibitem{FANTOM}  B. von Freyberg, T. Schaumann, W. Braun,
                  FANTOM User's Manual and Instructions: ETH Z\"urich,
                  Z\"urich, 1993 ;
                  B. von Freyberg, W. Braun,
                  {\it J. Comp. Chem.} {\bf 14}, 510 (1993).

\bibitem{HaOk99}  U.H.E. Hansmann, Y. Okamoto,
                    {\it J. Chem. Phys.} {\bf 110}, 1267 (1999).

\bibitem{Scholtz}  J.M. Scholtz, H. Qian, E.J. York, J.M. Stewart,
                  R.L. Baldwin,
                 {\it Biopolymers} {\bf 31}, 1463 (1991).

\end{thebibliography}
\end{document}